# Construction of Ito model for Geoelectrical Signals

Zbigniew Czechowski

Institute of Geophysics Polish Academy of Sciences, 01-452 Warsaw, Ks. Janusza 64, Poland, zczech@igf.edu.pl

Luciano Telesca*

National Research Council, Institute of Methodologies for Environmental Analysis, C.da S.Loja, 85050 Tito (PZ), Italy, luciano.telesca@imaa.cnr.it

**Abstract**: Ito stochastic differential equation governs one-dimensional diffusive Markov process. Geoelectrical signals measured in seismic areas can be considered as the result of competitive and collective interactions among system elements. The Ito equation may constitute a good macroscopic model of such phenomenon in which microscopic interactions are adequately averaged. The present study shows how to construct Ito model for a geoelectrical time series measured in a seismic area of southern Italy. Our results reveal that Ito model describes quite well the whole time series, but performs better when one considers fragments of the data set with lower variability range (absent or rare large fluctuations) . Our findings show that generally detrended geoelectrical time series can be considered as an approximation of the Markov diffusion process.

* Corresponding author: tel. +39-0971-427277, fax. +39-0971-427271, email luciano.telesca@imaa.cnr.it

# 1. INTRODUCTION

It is well known that the most relevant phenomenon that could originate the geoelectrical field is known as electrofiltration or streaming potential: the electrical signal is produced, when a fluid flows in a porous rock due to a pore pressure gradient. The phenomenon is generated by the formation within the porous ducts of a double electrical layer between the bounds of the solid, that absorbs electrolytic anions and cations distributed in a diffused layer near the boards. Due to pressure gradient, the fluid flows and transports a part of the cations, giving on one side of the layer an excess of positive charges. This produces an induced electric field along the length of the duct and the associated potential differences (streaming potential). The streaming potential can be responsible for the voltage measures on the ground surface preceding an earthquake[1, 2]. In a seismic focal region the increasing accumulation of strain can cause dilatancy of rocks [3]. The phenomenon of dilatancy consists in the formation and propagation of cracks inside a rock as stress reaches a critical value [4]. If the rocks in the focal region and surrounding volumes are saturated with fluids, the voids generate pressure gradients. Hence, fluids invade the newly opened voids and flow until the pressure balances inside the whole system of interconnected pores. During the fluid invasion the condition of rock hardening can be reached: the rock suddenly weakens and the earthquake is triggered. Therefore the structure of the geoelectrical signal is linked to the structure of the seismic focal zone.

It is also evident that geoelectrical signals can be considered as the result of collective or competitive interactions among system elements. Therefore, the construction of adequate macroscopic models (stochastic processes) describing observable phenomena may help in understanding the intrinsic processes on the microscopic level [5]. Up to now, the quantitative characterization of geoelectrical data measured in seismic areas was performed by using statistical tools, which were able to extract robust features hidden in their complex fluctuations. Fractal methods revealed the presence of scaling behaviour in earthquake-related geoelectrical signals, which are not realization of white noise process [6]. Appearance of the flicker-noise signatures and increase of the time series fractal dimension has been revealed in ULF (Ultra Low Frequency) geomagnetic signals before earthquakes [7-9]. Correlation between the power-law spectral exponent of geoelectrical signals and the Hurst exponent of seismic activity has been observed [6]. Long-range correlations in seismic electric signals preceding earthquakes have been studied using the rescaled range Hurst and detrended fluctuation analysis [10-13]. Such dynamics seem to be in agreement with the crack formation process and with the physical model of earthquake precursors due to crack propagation and charge dislocation model [14]. Multifractality was performed to get into insight the complexity of this type of signals, connected with the heterogeneity of the crust; the enhancement of the multifractal degree before the occurrence of an earthquake was related with the geometry and the structure of individual fault zones, that can be represented by a network with an anisotropic distribution of fracture orientations, and consisting of fault-related structures including small faults, fractures, veins and folds [15, 16]. Information-based statistical measure, like the Fisher Information Measure (FIM) and the Shannon entropy as well as the Entropy in natural time and the relevant complexity measures [28-30], were used in studying geoelectrical phenomena, revealing their ability in describing the complexity of the system [17-19] and suggesting their use as a reliable precursor of critical events [20-22].

In all the previous studies statistical methods are applied to quantitatively determine several parameters, able to describe the dynamics of the signals. What is still lacking concerns the construction of a macroscopic stochastic model of geoelectrical time series.

Usually time series may be considered as a realization of some stochastic process. If the process is diffusion and Markov (of order 1), the time evolution of the process can be described by the adequate Ito equation. There are methods of verification of the Markov property for the given time series [23] but it is very difficult to check if it is the diffusion one. Therefore, in the paper we assume that the time series under investigation fulfills these properties; the verification will be conducted by the comparison of the input time series with that generated by the reconstructed Ito equation.

Without a knowledge concerning details of the complex phenomenon and having time series data only, the Ito equation can compose a useful macroscopic model of the process.

The method of construction of the adequate Ito equation from time series data was proposed by Siegert et. al. [24]. This direct procedure, based on a histogram of the joined distribution function, leads to some approximation (clouds of points) of parameters in the Ito equation. The histogram method of modeling phenomena (which generate time series) fills a gap between linear stochastic models (ARMA etc.) and nonlinear deterministic models (which lead to a deteministic chaos; Takens embedding method [25]) because it is both stochastic and nonlinear. Moreover, Ito models give various types of distribution functions: from short-tail to long-tail distributions (see [26]). It should be underlined that unlike neural network models, which only reproduce data, the Ito model has the clear physical interpretation.

## 2. ITO MODEL

One-dimensional Ito equation

$$dy = a(y)dt + \sqrt{b(y)}dW(t) \qquad (1)$$

governs the evolution of the scalar diffusive Markov process $Y(t)$. Functions $a(y)$ and $b(y)$ are known to be drift and diffusion coefficients, respectively, and $W(t)$ is the Wiener process. The Ito equation is associated with the Fokker-Planck equation

$$\frac{\partial}{\partial t} p(y,t) = -\frac{\partial}{\partial y}[a(y)p(y,t)] + \frac{1}{2}\frac{\partial^2}{\partial y^2}[b(y)p(y,t)], \qquad (2)$$

which describes the behavior of the distribution function $p(y, t)$ of the process $Y(t)$. For the diffusion Markov process the short-term transition probability $P(y, t+\tau | y',t)$ has a Gaussian form

$$P(y,t+\tau | y',t) = \frac{1}{2\sqrt{\pi b(y')\tau}} \exp\left(-\frac{(y - y' - a(y')\tau)^2}{2b(y')\tau}\right). \qquad (3)$$

Therefore, knowing functions $a(y)$ and $b(y)$ we dispose (if we can solve Eqs. 1 and 2) of the full description of the stochastic process $Y(t)$.

Moreover, physical interpretation of the Ito equation is simple: it is the modified diffusion in the potential niche. The shape of the potential $V(y)$ is determined by the drift force; $V'(y) = -a(y)$. Strength of diffusive fluctuations $\sqrt{b(y)}dW(t)$ can be dependent on the current value of $Y(t)$.

However, the Ito equation constitutes only the macroscopic description of the phenomenon. Can we extract from the form of $a(y)$ and $b(y)$ any conclusion concerning microscopic aspects of the process – this is the important and difficult question. In the case of Brownian particle the interpretation of the linear force, $a(y) = -ay$, and the constant diffusion coefficient $b(y) = b$ is simple. However, the drift force need not to be connected to any external force (e.g. the drag force for the Brownian particle) and the diffusion term can be more complex than the random white noise. Both terms can be the result of the averaged interactive dynamics on the microscopic level. These effects were investigated for cases of toy models in [26] and [5].

## 3. DATA ANALYSIS

We analysed a geoelectrical record measured at Giuliano station (40.688°N, 15.789°E), located in southern Italy. The signal consists of voltage difference between two unpolarizable electrodes inserted 1m depth in the ground to avoid external temperature effects. The distance between the electrodes is 100m. Fig. 1 shows the analysed time series, which consists of minute-sampled geoelectrical values, recorded from March 1, 2001 and December 3, 2001.

In order to apply the Ito-based equations the signal was firstly reduced to a zero-mean stationary signal. Therefore the trend was removed by means of a two-sided moving average method, by using the following formula

$$x_k = \frac{1}{2s+1}\sum_{i=-s}^{s} x'_{k-i} \qquad (4)$$

with $s = 50$. The signal shows a daily fluctuation (Fig. 2), which was removed by means of method of differencing,

$$y(i) = x(i) - x(i\text{-}1440). \qquad (5)$$

The data detrended and deseasonalized are shown in Fig. 3.

The histogram method [24] was used to reconstruct the functions $a(y)$ and $b(y)$ in Ito equations from the time series by means of the equations [27]:

$$a[y(t)] = \lim_{\tau \to 0} \int_{-\infty}^{\infty} \frac{1}{\tau}[\tilde{y}(t+\tau) - y]p(\tilde{y}, t+\tau|y,t)d\tilde{y}, \qquad (6)$$

$$b[y(t)] = \lim_{\tau \to 0} \int_{-\infty}^{\infty} \frac{1}{\tau}[\tilde{y}(t+\tau) - y][\tilde{y}(t+\tau) - y]p(\tilde{y}, t+\tau|y,t)d\tilde{y}, \qquad (7)$$

where $\tilde{y}(y+\tau)$ is the solution of the Ito equation after time $\tau$ (when the initial condition at time t is $\tilde{y}(t) = y(t)$) and $p(\tilde{y}, t+\tau|y,t)$ is the conditional distribution function. The distribution function is approximated by using histograms of joint distribution function $p(\tilde{y}, t+\tau; y,t)$ and of marginal distribution function $p(y,t)$ according to the formula

$$p(\tilde{y}, t+\tau|y,t) = \frac{p(\tilde{y}, t+\tau; y,t)}{p(y,t)}. \qquad (8)$$

2   In the histogram method we replace the integrals in Eqs.6 and 7 by sums and we omit

3   limits. The time increment $\tau$ represents the time step in time series.

4   We analysed: i) the whole data series; ii) one subseries of relatively large variability

5   (subseries 1: from datum 1 to datum $10^5$); and iii) one subseries of relatively small

6   variability (subseries 2: from datum 195000 and datum 220000). The analysis of the

7   two subseries is motivated by the investigation of variation of the Ito model for the

8   geoelectrical series with the variability range.

9   The histograms for the joined distribution function $p(y_t, y_{t+1})$ of the whole data set as

10  well as the two subseries are shown in Fig. 4. In two cases (the whole data and the

11  subseries 1) the variability range of $y_t$ was divided in 150 small equal sectors

12  (partitions), thus 150x150 histogram columns $p(i, j)$ were calculated. The histogram

13  of the subseries 2 (only 60x60 columns to keep the same grid size, because variability

14  range of $y_t$ is smaller) has a clear flattening, which reveals stronger correlations

15  (weaker randomness) respect to the whole series and the subseries 1.

16  Stationary marginal distribution functions $p(y)$ are shown in Fig.5a (in lin-log

17  scales) and in Fig. 5b (in log-log scales). Two different behaviors can be viewed for

18  the whole series (red symbols): exponential for $y < 0.5$ and inverse-power for $1.8 > y$

19  $> 0.5$ (see Table 1 in [26]). The excess of large fluctuations is evidenced by a hump

20  in Fig. 5b for $p(y)$, where $y$ is between 2 and 5.

21  Similar shape is shown by subseries 1, where the excess of large fluctuation is

22  evidenced by a hump around $y = 4$. Different shape is revealed by subseries 2, where

23  the deviation from the exponential form is relatively small for $y > 0.5$.

All these forms of $p(y)$ are caused by two competitive forces in the Ito model: the drift $a(y)$ and the stochastic force $\sqrt{b(y)}dW(t)$. By using the histograms $p(i, j)$ and $p(i)$ the functions $a(y)$ and $b(y)$ can be deduced (see Figs. 6 and 7) according to Eqs. 6 and 7.

It should be underlined that for more dispersed regions ($|y| > 1$) the fitting is very uncertain. An introduction of the fast decrease of the function $a(y)$ and jumps of $b(y)$ for $|y| > 3$ was necessary to prevent too big (infinite) fluctuations of $y$.

We observe that for $|y| < 0.7$ both functions have a similar behavior for the three cases under investigation: the linear decreasing function $a(y)$, and $b(y)$ approximately of the form $b(y) = |y|$ ), which leads to exponential forms (see Table 1 in [26]) of $p(y)$ in this range. Greater fluctuations in the time series have a different origin, revealed by the different behavior of functions $a(y)$ and $b(y)$ for $|y| > 0.7$. A good illustration of different microscopic mechanisms is a shape of the potential niche (Fig. 8) which becomes much wider for $|y| > 0.7$ allowing for greater fluctuations and, therefore, long tails (for whole series and subseries 1) of the distribution function $p(y)$. It should be underlined that in the case of the whole data series the functions $a(y)$ and $b(y)$ comprise large fluctuations in all regions of time series and, therefore, give some mean results for them.

The Ito equation represents the macroscopic model of the phenomenon; it describes the time evolution of the variable which is observable. However, some limited information about the influence of the microscopic mechanisms may be deduced. The exponential small fluctuations differ (they have longer tail) from the purely random gaussian noise (which is typical in many time series) and they are the result of the $y$-dependent diffusion coefficient (i.e. the function $b(y)$) – greater states $y$ experience longer random jumps. Larger inverse-power fluctuations correspond to other intrinsic mechanism, which is the cause of the wider potential niche (and some changes in $b(y)$). The above mentioned effects are included in the form (Eq. 3) of the short-term transition probability. In order to understand and explain this macroscopic behavior we should investigate in details the microscopic physical nature of the complex phenomenon, however, this is beyond the scope of this paper.

On the other hand, to this aim, there were derived Ito equations for cases of toy models, in which the whole microscopic evolution is controlled [5, 26].

A comparison of the input detrended and deseasonalized time series with the time series generated by the constructed Ito equation is presented in Figs. 9, 10 and 11. The original time series and the Ito-model reconstructed time series are quite similar, even if the original time series in the whole case and in the subseries 1 shows a certain spiky character of the large fluctuations not visible in the series generated by the reconstructed Ito equations. However, this is a result of a stochastic character of the Ito equations. The Ito-based reconstructed time series for the subseries 2, instead, appears very similar to the original time series.

## Conclusions

A stochastic modeling always requires the ergodicity of the time series under investigation, in order to estimate the stochastic properties from these data. Therefore, we had to remove the trend and seasonal fluctuations. However, the obtained time series appeared to be not stationary as the whole; there are stationary fragments. Constructed Ito models quite well describe the fragments, it seems that the detrended and deseasonalized geoelectrical time series can be considered as an approximation of the Markov diffusion process. However, in the time series generated by Ito equations great fluctuations are spontaneous due to stochastic character of models. On the other hand, some large fluctuations can result ("fingerprints" of the trend) from the trend (which may have a deterministic origin); it could be an explanation of the assembling of great fluctuations together - seen in the input time series. The constructed short-term transition probability can be a useful tool in a short time prediction.

**Acknowledgementes**

The authors are grateful to two anonymous referees. The study presented in this paper was supported by the CNR-PAN Bilateral Project "Development of innovative nonlinear time series tools for dynamical systems" in the framework of the Bilateral Agreement for Scientific and Technological Cooperation between CNR and PAN, 2010-2012.

**Figure Captions**

Fig.1. Geoelectrical data

Fig.2. Excerpt of the original data, where the daily oscillation is visible

Fig.3. Detrended and deseasonalized data from Fig.1.

Fig.4. Histograms of the joined distribution function for a) the whole data set, b) subseries 1, c) subseries 2.

Fig.5. Comparison of marginal distribution functions $p(y)$. Red symbols represent the whole series, blue ones the subseries 1 and green ones the subseries 2: a)in lin-log scale, lines are graphs of exponential functions: ~ $Exp[-13x]$ (green), ~ $Exp[-7.4x]$ (red), ~$Exp[-6.8x]$ (blue) ; b)in log-log scale, lines are graphs of inverse-power functions: ~ $x-4$ (green), ~ $x-3.5$ (red), ~ $x-2.6$ (blue)

Fig.6. Comparison of functions $a(y)$. Red symbols represent the whole series, blue ones the subseries 1 and green ones the subseries 2.

Fig.7. Comparison of functions $b(y)$. Red symbols represent the whole series, blue ones the subseries 1 and green ones the subseries 2.

Fig.8 Comparison of potentials $V(y)$. Red line represent the whole series, blue one the subseries 1 and green one the subseries 2.

Fig.9 (a). The input detrended and deseasonalized whole time( 400000 data ). (b). The time series generated by the series reconstructed Ito equation.

Fig. 10. (a).The input detrended and deseasonalized time Subseries 1. (b). The time series generated by the reconstructed Ito equation.

Fig. 11. (a). The input detrended and deseasonalized time Subseries 2. (b). The time series generated by the reconstructed Ito equation.

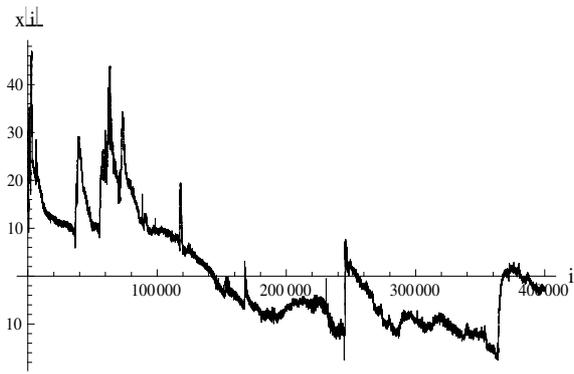

Fig.1. Geoelectrical data

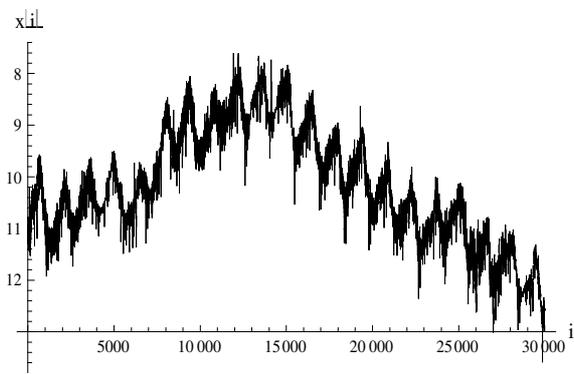

Fig.2. Excerpt of the original data, where the daily oscillation is visible

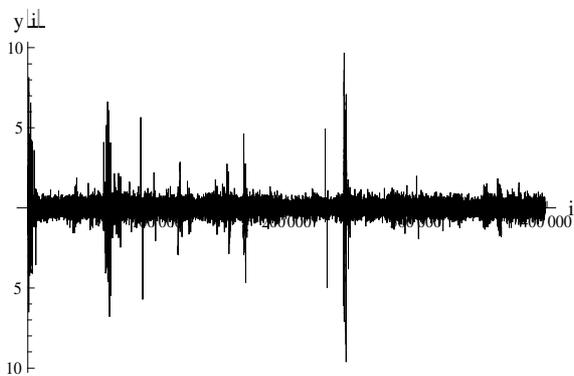

Fig.3. Detrended and deseasonalized data from Fig.1.

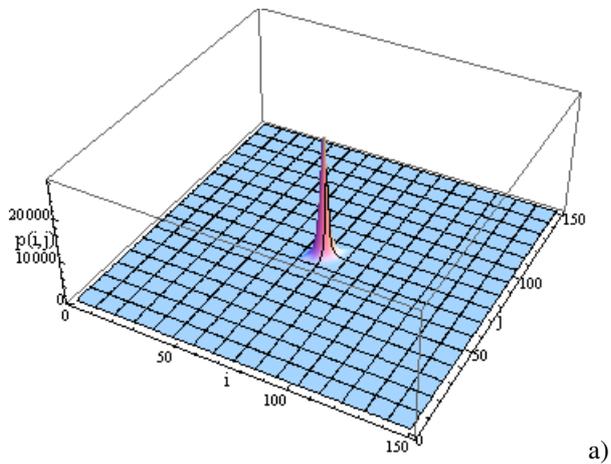

a)

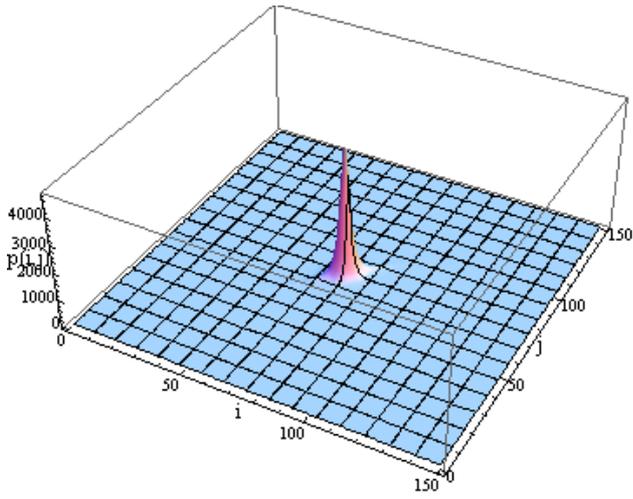

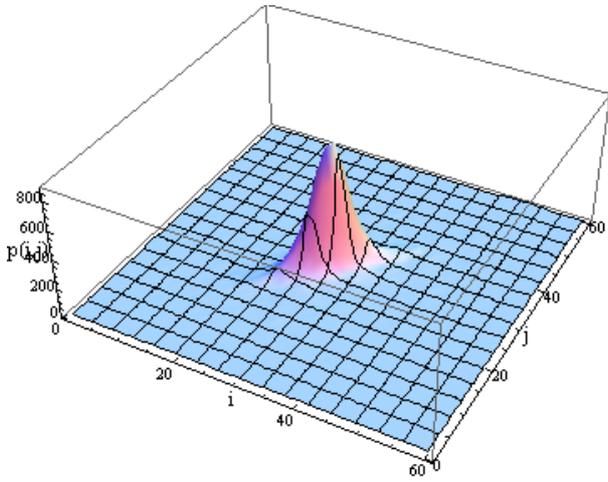

c)

Fig.4. Histograms of the joined distribution function for a) the whole data set, b) subseries 1, c) subseries 2.

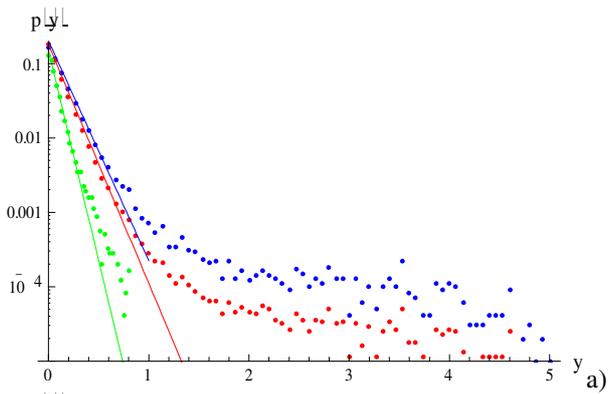

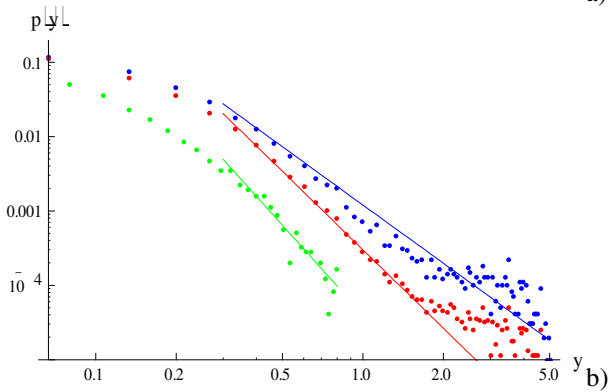

Fig.5. Comparison of marginal distribution functions $p(y)$. Red symbols represent the whole series, blue ones the subseries 1 and green ones the subseries 2:
- (a) in lin-log scale, lines are graphs of exponential functions: ~ Exp[-13x] (green), ~ Exp[-7.4x] (red), ~Exp[-6.8x] (blue)
- (b) in log-log scale, lines are graphs of inverse-power functions: ~ $x^{-4}$ (green), ~ $x^{-3.5}$ (red), ~ $x^{-2.6}$ (blue)

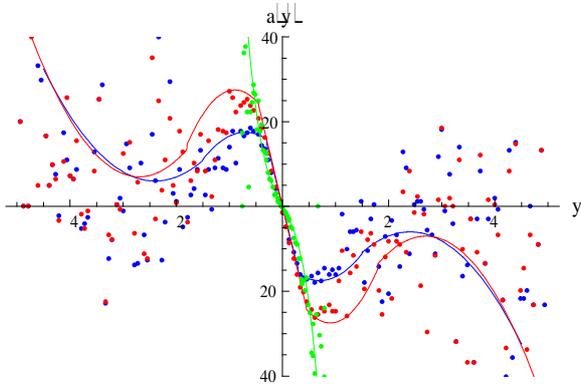

Fig.6. Comparison of functions $a(y)$. Red symbols represent the whole series, blue ones the subseries 1 and green ones the subseries 2.

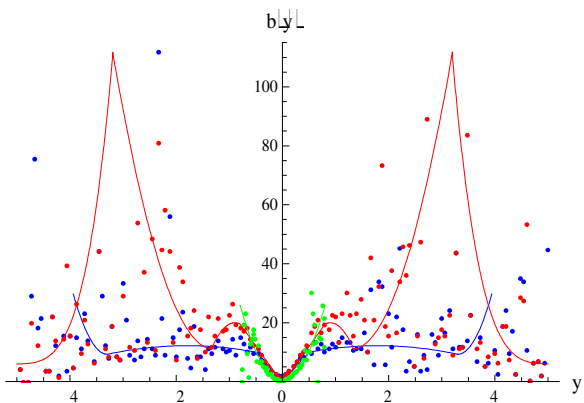

Fig.7. Comparison of functions $b(y)$. Red symbols represent the whole series, blue ones the subseries 1 and green ones the subseries 2.

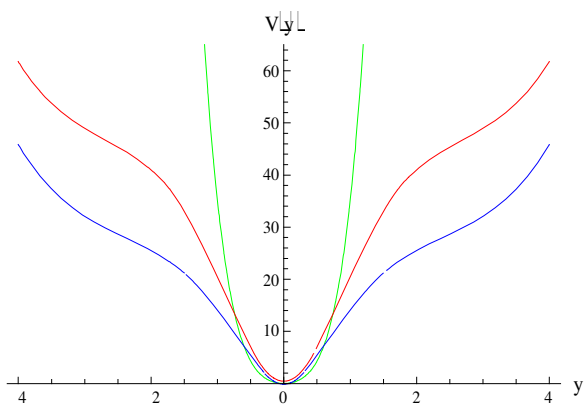

Fig.8 Comparison of potentials $V(y)$. Red line represent the whole series, blue one the subseries 1 and green one the subseries 2.

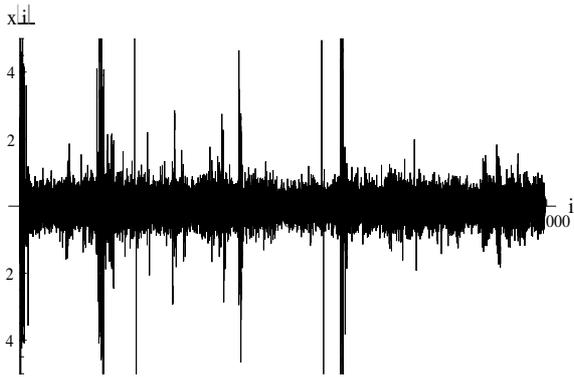

Fig.9a. The input detrended and deseasonalized whole time series ( 400000 data )

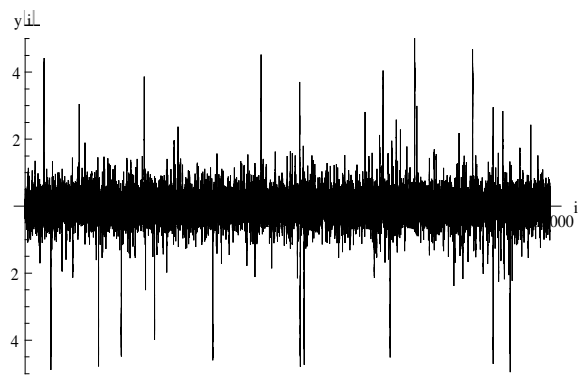

Fig.9b. The time series generated by the reconstructed Ito equation

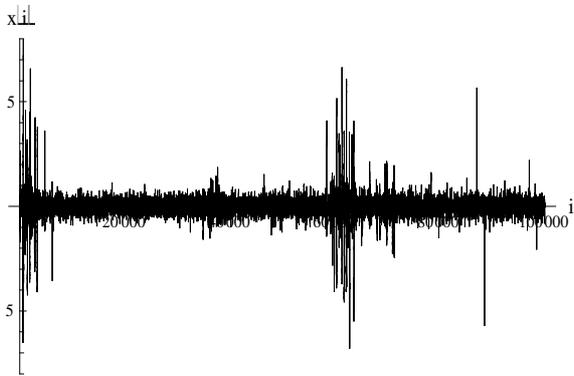

Fig.10a. The input detrended and deseasonalized time Subseries 1

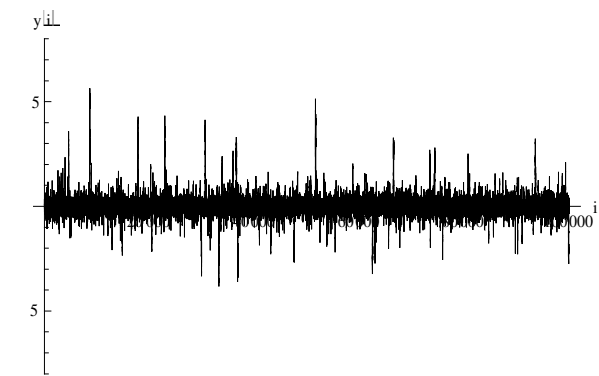

Fig.10b. The time series generated by the reconstructed Ito equation

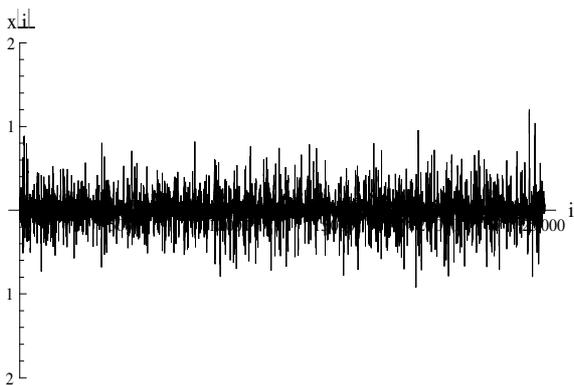

Fig.11a. The input detrended and deseasonalized time Subseries 2

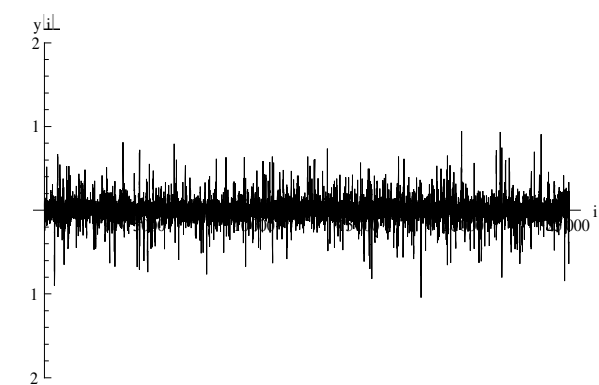

Fig.11b. The time series generated by the reconstructed Ito equation